\documentclass[%
twocolumn,
superscriptaddress,
showpacs,
preprintnumbers,
graphicx,
floatfix,
byrevtex,
amsfonts,
amsmath,
amssymb,
bibnotes,
apl,
]{revtex4-1}

\usepackage{float,graphicx}
\usepackage{dcolumn}
\usepackage{bm}
\usepackage[mathlines]{lineno}
\usepackage{ulem}
\usepackage{color}
\usepackage[version=3]{mhchem}
\usepackage{mathrsfs,braket}

\newcommand*\tmote{1T'$-\mathrm{MoTe_2}$\ }
\newcommand*\tdmote{$\mathrm{T_d-MoTe_2}$\ }

\begin{document}
\preprint{} 

\title[Short Title]{Ultrafast dynamics of the low frequency shear phonon in 1T'$-$\ce{MoTe2}}

\author{Takumi Fukuda}
\email{s1920338@s.tsukuba.ac.jp}
\affiliation{Department of Applied Physics, Graduate school of Pure and Applied Sciences, University of Tsukuba, 1-1-1 Tennodai, Tsukuba 305-8573, Japan}
\author{Kotaro Makino}
\affiliation{Nanoelectronics Research Institute, National Institute of Advanced Industrial Science and Technology (AIST), Tsukuba Central 5, 1-1-1 Higashi, Tsukuba 305-8565, Japan}
\author{Yuta Saito}
\affiliation{Nanoelectronics Research Institute, National Institute of Advanced Industrial Science and Technology (AIST), Tsukuba Central 5, 1-1-1 Higashi, Tsukuba 305-8565, Japan}
\author{Paul Fons}
\affiliation{Nanoelectronics Research Institute, National Institute of Advanced Industrial Science and Technology (AIST), Tsukuba Central 5, 1-1-1 Higashi, Tsukuba 305-8565, Japan}
\author{Alexander V. Kolobov}
\affiliation{Nanoelectronics Research Institute, National Institute of Advanced Industrial Science and Technology (AIST), Tsukuba Central 5, 1-1-1 Higashi, Tsukuba 305-8565, Japan}
\affiliation{Department of Physical Electronics, Faculty of Physics, Herzen State Pedagogical University of Russia, 48 Moika Embankment, St. Petersburg, 191186, Russia}
\author{Keiji Ueno}
\affiliation{Graduate School of Science and Engineering, Saitama University, Saitama 338-8570, Japan}
\author{Muneaki Hase}
\email{mhase@bk.tsukuba.ac.jp}
\affiliation{Department of Applied Physics, Graduate school of Pure and Applied Sciences, University of Tsukuba, 1-1-1 Tennodai, Tsukuba 305-8573, Japan}


\begin{abstract}
We report on the dynamics of coherent phonons in semimetal 1T'$-$\ce{MoTe2} using femtosecond pump$-$probe spectroscopy. On an ultrafast sub$-$picosecond time scale at room temperature, a low frequency and long$-$lifetime shear phonon mode was observed at 0.39 THz, which was previously reported in the form of a characteristic phonon only in the low temperature \tdmote phase.
Unlike the other optical phonon modes, the shear phonon mode was found to strongly couple with photoexcited carriers. Moreover, the amplitude of the shear mode surprisingly decreased with increasing excitation density, a phenomenon which can be attributed to be a consequence of the lattice temperature rising after excitation.
These results provide useful physical information on ultrafast lattice symmetry switching between the normal semimetal 1T' and the lattice inversion symmetry breaking Type$-$II Weyl semimetal T$_\mathrm{d}$ phases.

\end{abstract}

\maketitle

\begin{figure*}
	\centering
	\includegraphics[width=12cm]{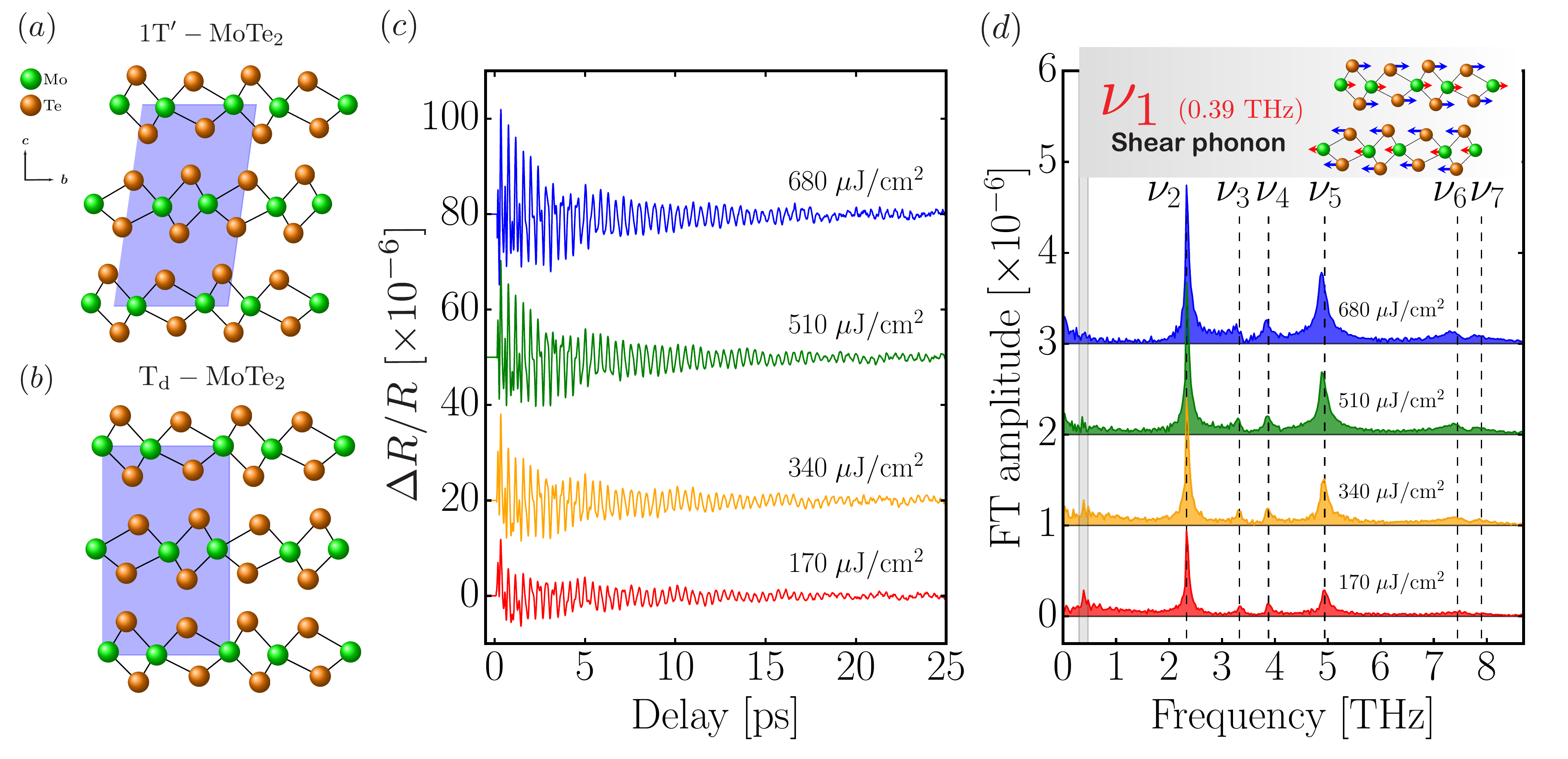}
	\caption{($a$) and ($b$) are the crystal structures of monoclinic \tmote and orthorhombic \tdmote, respectively. The blue colored areas represent the unit cells. ($c$) Fluence dependent  (170 to 680 $\mu\mathrm{J/cm^2}$) time domain signals of CPs in 1T' phase obtained from the $\Delta R/R$ signal, where the excited carrier response has been subtracted as background. ($d$) FT amplitude spectra of ($c$) and a schematic image of the shear phonon of the $\nu_1$ mode in inset. 
	}
	\label{results}
\end{figure*}

Recent interest in transition metal dichalcogenides (TMDCs) has dramatically increased owing to the variety of unusual properties stemming from the existence of two dimensional Van der Waals (VdW) structures, such as graphene \cite{novoselov2004electric,novoselov2005twodimgasdirfer,geim2009graphene}.
One of the reasons that drives research in TMDCs
is that they intrinsically possess a wide variety of possible crystal and electronic structures depending on the large number of accessible chemical combinations between metal $(\ce{Mo},\ce{W},\cdots)$ and chalchogen $(\ce{S},\ce{Se},\ce{Te})$ atoms, as well as the accessibility of other phases by
low energy photon excitation \cite{PhysRevB.94.094114}.
These versatile properties can be useful for further understanding of the fundamental properties of topological materials as well as the advancement of high-performance and functional electronic, optoelectronic and quantum devices \cite{excitonflux,wang2012electronics,chhowalla2013chemistry},
particularly the next generation of strain$-$engineered phase$-$change materials \cite{song2015room} or phase patterning technology driven by light \cite{cho2015phase}.
Within the TMDC family, \ce{MoTe2} has attracted considerable attention because of the presence of rich structural and electronic phases. \ce{MoTe2} has three possible structural phases, the trigonal prismatic 2H semiconducting phase (the most stable at room temperature: RT), the centrosymmetric monoclinic 1T' semimetal phase (metastable at RT), and the lattice symmetry breaking orthorhombic T$_\mathrm{d}$ semimetal phase (stable below 250 K). Among these phases, the 1T' and T$_\mathrm{d}$ structures are similar, except for distortions along the $a-$ or $b-$axes as shown in Fig.\ref{results}($a$) and ($b$).
\ce{MoTe2} is also known for being a candidate topological phase material such as a topological insulator (TI) or a Weyl semimetal (WSM) due to the breaking of lattice inversion symmetry by certain layer configurations.
First-principle calculations and observations of Fermi arcs in lattice symmetry breaking orthohombic T$_\mathrm{d}-$\ce{MoTe2} and similar structure, T$_\mathrm{d}-$\ce{WTe2}, have led to the establishment and confirmation of the theory of Type$-$II WSMs with intriguing carrier transport properties with unusually long lifetimes \cite{soluyanov2015type,sun2015prediction,huang2016spectroscopic,deng2016experimental, crepaldi2017enhanced}.\par
Very recently, some groups have observed the ultrafast dynamics of coherent phonons (CPs) and lattice symmetry switching between the 1T' and $\mathrm{T_d}$ phases, which corresponds to a phase transition from the WSM to the normal phases.
Sie \textit{et al}. induced shear displacements using THz pump pulses in the T$_\mathrm{d}$ phase of \ce{WTe2}, which was monitored by ultrafast electron diffraction \cite{sie2019ultrafast}. Zhang \textit{et al.},
on the other hand, observed a similar phenomenon using optical pump$-$probe reflectivity measurements from extremely low temperature 4K to near RT \cite{zhang2019light}, corresponding to the low temperature T$_\mathrm{d}$ phase. However, the related CPs at RT of the \tmote phase have not been well investigated.
Therefore it is still unclear if photoexcitation of the high temperature 1T' phase induces a phase transition in \ce{MoTe2}. It is important to gain insight into the photoexcited state at RT and even higher temperatures in TMDCs, because of potential device applications in general temperature conditions.
\par

Here, we investigate the dynamics of CPs in monoclinic \tmote at RT using a femtosecond optical pump$-$probe reflectivity technique with a high signal to noise ratio. %
In addition to the typical Raman$-$active optical modes of the 1T' phase, %
a $\mathrm{T_d}-$originated shear vibration mode is also clearly observed in the high temperature 1T' phase.
The time$-$domain CP oscillations are analyzed using a short window Fourier transform method, which enables the extraction of the dynamics related to the generation and decay of each phonon mode in time$-$frequency space.
Our results indicate that the light$-$induced structural phase transition between the adjacent 1T' and T$_\mathrm{d}$ phases, here referred to as "lattice symmetry switching", occurs even at RT. We suggest that the ultrafast dynamics of such switching is an important research subject necessary to explore the applications of \ce{MoTe2} as a Type$-$II WSM.

Optical pump$-$probe measurements were carried out using a Ti:sapphire oscillator operated at 80 MHz, which provided near infrared optical pulses with a pulse duration of $\leq$ 30 fs and a central wavelength of 830 nm (with a corresponding photon energy of $\hbar \omega \sim 1.5\ \mathrm{eV}$). It should be noted that the continuous heating effect by the $80\ \mathrm{MHz}$ repetition laser is negligibly small, as demonstrated by the phonon spectra shown below exhibiting the almost same peak positions as those of previous Raman studies on bulk \tmote.
The average fluence of the pump beam was varied from $F = 170$ to 680 $\mu\mathrm{J/cm}^{2}$. The upper limit of fluence (680 $\mu \mathrm{J/cm^2}$) was set to prevent surface damage.
The $s-$polarized pump and the $p-$polarized probe beams were co$-$focused onto the sample to a spot size of $\approx$\ 8 $\mu$m with an incident angles of about $5^{\circ}$ and $15^{\circ}$ with respect to the sample normal, respectively.
The optical penetration depth at 830 nm was estimated to be 50 nm from the complex dielectric function as calculated by density functional theory (DFT) simulations.
The delay between the pump and probe pulses was scanned up to 30 ps \cite{hase2012frequency} by an oscillating retroreflector operated at a frequency of 9.5 Hz.
The transient reflectivity change ($\Delta R/R$) was recorded as a function of pump$-$probe time delay. The measurements were performed in air at RT. The figures used in this article were generated using Matplotlib \cite{Hunter:2007}.

The sample used was a small flake of a \tmote single crystal with the $c-$axis of the crystal corresponding to the sample normal. The \tmote bulk crystal was prepared by Chemical Vapor Transport \cite{ueno2015introduction}. 
The crystal thickness was $\geq$ 500 $\mu$m and strain and/or confinement effects were negligible \cite{he2018PRB}. In addition, Raman measurements on our \tmote sample indicated no peak at $13\ \mathrm{cm^{-1}} (= 0.39\ \mathrm{THz})$ and thus confirming that our sample at RT does not include the $\mathrm{T_d}$ phase.


Fig.\ref{results}($c$) shows the $\Delta R/R$ signals measured for the \tmote phase
with sub$-$picosecond time resolution
for a variety of pump fluences ranging from 170 to 680 $\mu\mathrm{J/cm}^{2}$. CP oscillations can be clearly observed up to $\sim$ 25 ps time delay at RT.
The oscillations are not simply sinusoidal, suggesting the existence of multiple phonon modes driven by charge$-$density fluctuations, which traditionally described by resonant impulsive stimulated Raman scattering (ISRS) \cite{garrett1996coherent,PhysRevB.65.144304} or by a displacive excitation of CP (DECP) mechanism \cite{zeiger1992theory}. Under the condition that the pump photon is absorbed by an opaque sample, the driving force can be described by the Raman tensor, which includes both the resonant ISRS and DECP mechanisms. In the following we will briefly introduce the DECP mechanism.
DECP theory describes that the initial phonon coordinate $Q_0(t)$ is proportional to the photoexcited carrier density $n(t)$, namely,
\begin{equation}
	Q_0(t) = \kappa n(t),
	\label{DECP}
\end{equation}
where $\kappa$ is a constant. Considering the relation $\Delta R(t)/R \propto Q_0$, the time evolution of phonon amplitude of CP generated by the DECP process should follow the dynamics of photoexcited carrier density $n(t)$.

Fig.\ref{results}($d$) shows the Fourier transformed (FT) spectra of the time domain CP signals for \tmote obtained for different pump fluences. According to the FT spectra, it can be seen that a total of seven different phonon modes can be observed in the $\Delta R/R$ data.
Each of the phonon frequencies has been labeled as $\nu_1 \sim \nu_7$ from the lowest to the highest frequency. In detail, $\nu_1 = 0.39 \ \mathrm{THz}$,  $\nu_2 = 2.34\ \mathrm{THz}$, $\nu_3 = 3.34 \ \mathrm{THz}$, $\nu_4 = 3.88\ \mathrm{THz}$,
$\nu_5 = 4.94\ \mathrm{THz}$, $\nu_6 = 7.48\ \mathrm{THz}$ and $\nu_7 = 7.90\ \mathrm{THz}$.
In particular, $\nu_2 \sim \nu_7$ modes, which are observed by the present optical pump$-$probe method, are consistent with the Raman spectra of the 1T' phase at RT, as reported recently by various groups \cite{chen2016activation,zhang2016raman,lai2018anisotropic,ma2016raman}.

The low frequency $\nu_1$ mode is, however, absent in the Raman spectra of the 1T' phase at RT, but is characteristic of the $\mathrm{T_d}$ phase, which is a low temperature phase ($<250$K) of \ce{MoTe2}. Note that $\nu_1$ can be assigned to the $A_1$ mode in the $\mathrm{T_d}$ phase or as the $\mathrm{B_u}$ mode in the 1T' phase, while the $\nu_2 \sim \nu_7$ phonon modes can be assigned as the $A_1$ modes in the $\mathrm{T_d}$ phase or as the $A_g$ mode in the 1T' phase, according to previous reports \cite{chen2016activation,zhang2016raman,ma2016raman}.
It is generally impossible to observe an infrared$-$active $\mathrm{B_u}$ phonon mode in pump$-$probe measurements based on resonant ISRS or DECP processes \cite{garrett1996coherent,PhysRevB.65.144304,zeiger1992theory}.
This implies, in our experiments, the low frequency $\nu_1$ mode cannot be the $\mathrm{B_u}$ mode of the 1T' phase. Thus, the low frequency $\nu_1$ phonon observed in the present study is assigned to the $A_1$ mode originating from the $\mathrm{T_d}$ phase as discussed in detail below.

The low frequency $\nu_1$ mode can be considered to be an interlayer shear vibration mode\cite{zhang2019light,sie2019ultrafast}, a mode that is a key to lattice symmetry switching \cite{chen2016activation,zhang2016raman} as can be seen in the inset of Fig.\ref{results}($d$).
It can also be seen that the asymmetric shape of the FT amplitude around $\nu_4$ at low excitation density, as shown in Fig.\ref{results}($d$) at $F \leq$ 340 $\mathrm{\mu J/cm^2}$, implies the existence of a splitting of a phonon mode ($\mathrm{131\ cm^{-1}} = 3.93\ \mathrm{THz}$), which is also characteristic of the $\mathrm{T_d}$ phase. Thus, 
the fact that the low frequency $\nu_1$ mode was observed even at RT suggests the possibility of a transient structural transition to the $\mathrm{T_d}$ phase or a mixture of the two competing 1T' and $\mathrm{T_d}$ phases upon photoexcitation. 
It was also noted, in the present study, that the 2H phase does not play a role in the structural phase transition.

\begin{figure}
	\centering
	\includegraphics[width=8cm]{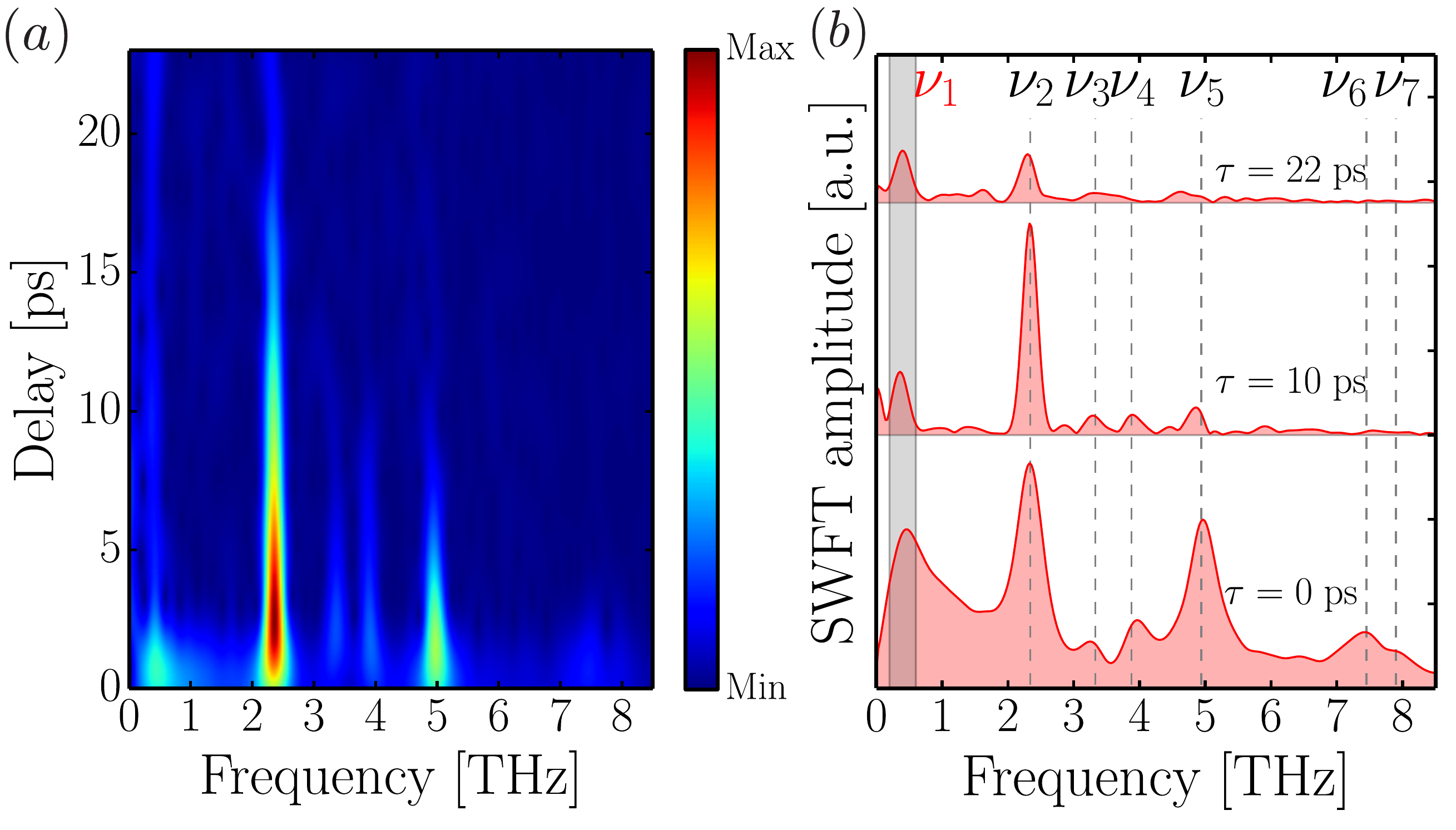}
	\caption{($a$) SWFT spectrogram of the $\Delta R/R$ signal for  $F=170\ \mu \mathrm{J/cm^2}$ of ($a$). The FT amplitude spectrum at $\tau = 0, 10, 22\ \mathrm{ps}$ (in ($a$)) are displayed in ($b$). 
	}
	\label{STFT1}
\end{figure}

Fig.\ref{STFT1}($a$) shows a short window FT (SWFT) spectrogram \cite{gambetta2006real} for $F$ = 170 $\mu$J/cm$^{2}$, a time$-$frequency$-$domain representation of Fig.\ref{results}($a$). 
Here, the FWHM of the Gaussian window function was set as 2.0\ $\mathrm{ps}$ to eliminate any artifacts due to edge effects while allowing the detection of the low ($<$ 0.5 THz) frequency $\nu_1$ mode.
The dominant optical phonon modes as well as the $\nu_1$ mode can be seen, and they decay with dramatically different time constants.
In particular, the lifetimes of the $\nu_3 \sim \nu_7$ modes were found to decay within several picoseconds, which is a typical behavior of CPs, however, the $\nu_1$ and $\nu_2$ modes persisted beyond $\approx 22\ \mathrm{ps}$ up to the limit of the time delay of the measurement. To explore this in more detail, we performed a fit of the time$-$domain data for $F = 170\ \mu\mathrm{J/cm^2}$ shown in Fig.\ref{results}($c$) using damped oscillations with the three dominant frequencies $\nu_1$, $\nu_2$ and $\nu_5$,
\begin{equation}
	\frac{\Delta R (t)}{R} = \sum_{i  = 1,2,5}\xi_i e^{-\frac{t}{\tau_i}}\cos(2\pi\nu_i t + \varphi_i),
	\label{FIT}
\end{equation}
where $\xi_i$ is the amplitude, $\tau_i$ is the relaxation time, $\nu_i$ is the frequency and $\varphi_i$ is the initial phase of the CP. The fit indicates $\tau_1 = 21.7 \pm 1.4\ \mathrm{ps}$, $\tau_2 = 9.8 \pm 0.04\ \mathrm{ps}$ and $\tau_5 = 3.7 \pm 0.1\ \mathrm{ps}$. The $\nu_1$ mode exhibits an even much longer lifetime than that of the $\nu_2$ and $\nu_5$ modes.
It is noted that these relaxation times only weakly depend on the pump fluence, and therefore we will not discuss this effect here.
The fit also indicates $\varphi_1 = 27 \pm 2^\circ$, $\varphi_2 = 90 \pm 1^\circ$, and $\varphi_5 = 96 \pm 1^\circ$. Considering the value of $\varphi_1$ obtained above, the initial phase of the $\nu_1$ mode is found to be cosine$-$like, while the other modes,  $\varphi_2$ and $\varphi_5\ (\approx \pi/2)$, exhibits sine$-$like behavior. 
It can also be seen that the shear vibration mode ($\nu_1$) exhibits a larger amplitude than that observed in static spectra (Fig.\ref{results}($d$)) just after the excitation by the pump pulse ($\tau = 0\ \mathrm{ps}$) as seen in Fig.\ref{STFT1}($b$).

\begin{figure}
	\centering
	\includegraphics[width=6cm]{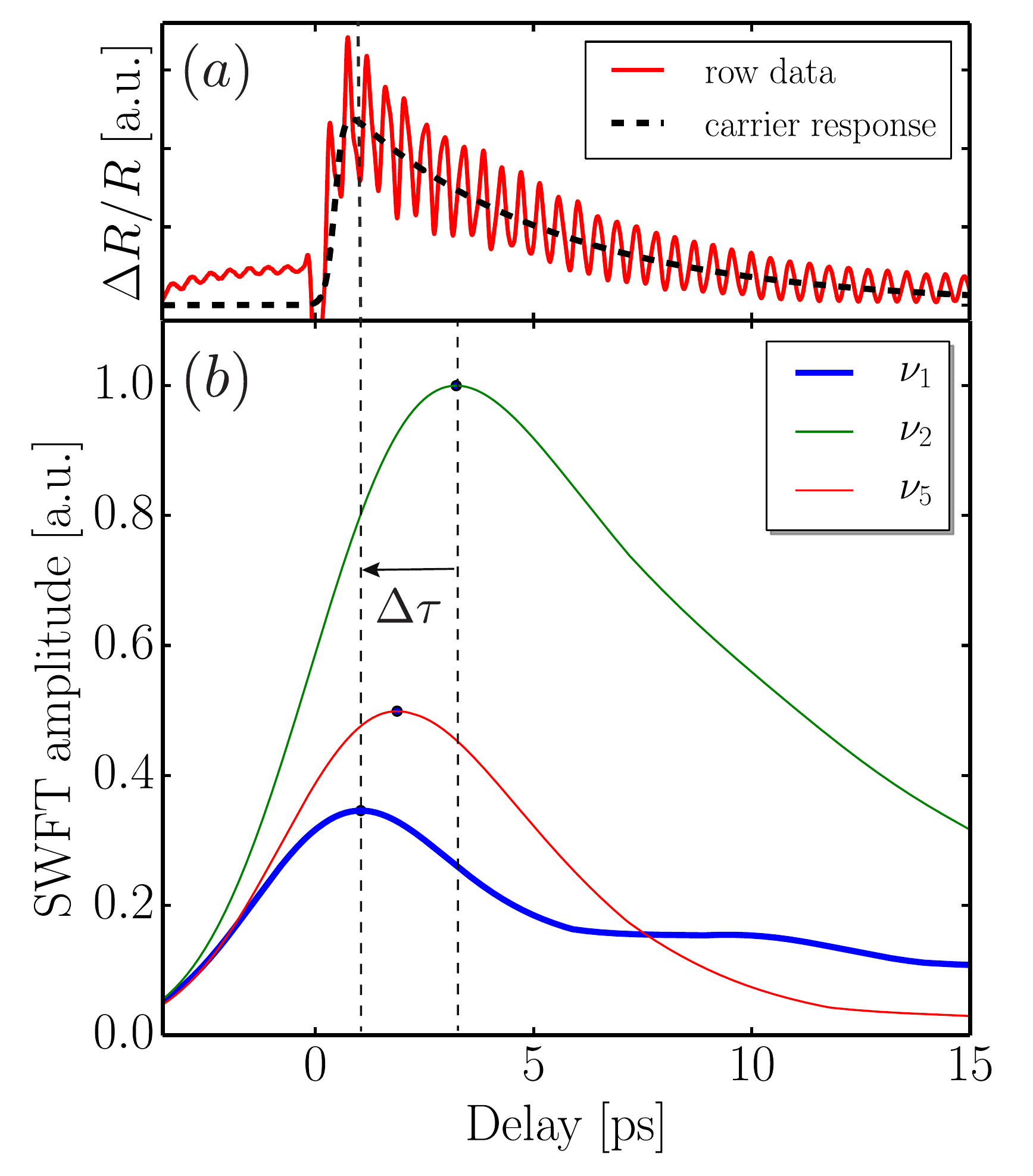}
	\caption{($a$) Part of $\Delta R/R$ signal including carrier relaxation and the phonon oscillations of 1T' phase at 170 $\mu \mathrm{J/cm^2}$. The dashed line represents the carrier relaxation in the signal. ($b$)Time evolution of the SWFT amplitudes for $\nu_1,\ \nu_2,\ \nu_5$ modes as a function of the time delay. The vertical dashed lines represent the peak positions. $\Delta \tau$ represents the time lag between the $\nu_1$ and $\nu_2$ modes.
	}
	\label{STFT2}
\end{figure}

Fig.\ref{STFT2} shows the time evolution of the peak amplitude for
the three dominant modes $\nu_1,\ \nu_2,$  and $\nu_5$ , shown in Fig.\ref{STFT1}($a$). According to Fig.\ref{STFT2}($b$), the rise time of the amplitude of the $\nu_1$ mode is the fastest among all the vibrational modes (this is also the case for the other optical modes although the data is not shown) just after excitation.
It was also found that the rise of the $\nu_1$ mode coincides with that of the photoexcited carrier dynamics shown in Fig.\ref{STFT2}($a$) in accordance with eq.(\ref{DECP}). In addition to the results in Fig.\ref{STFT2}, the fact that the $\nu_1$ mode exhibits displacive behavior (cosine$-$like), while the other modes exhibit impulsive behavior (sine$-$like), indicates the $\nu_1$ mode strongly couples with the photogenerated carriers and is excited preferentially over all the phonon modes.
Thus, under low excitation density (170\ $\mathrm{\mu J/cm^2}$) in which the lattice temperature rise is negligibly small, photoexcited carriers exert a step function electrostrictive force on the lattice driving the coherent $\nu_1$ mode, which induces the lattice symmetry of 1T' phase to partially change toward $\mathrm{T_d}$ phase.\par
Based on the relationship between the carrier (electron) relaxation time and the initial phase of the coherent phonons for simple semimetals (i.e., tan$\varphi = \Gamma/\Omega$, where $\Gamma$ is the reciprocal of carrier relaxation time and $\Omega=2\pi\nu$) \cite{zeiger1992theory,li2013optical}, Fig.\ref{STFT2} also suggests a different relaxation time as the driving force of three dominant phonon modes.  That is the shorter the carrier relaxation time $\tau_{\mathrm{el}}$, the stronger the impulsive behavior (sine$-$like). In fact, the observation that the displacive$-$like $\nu_1$ mode strongly couples with photoexcited carriers, indicates a longer relaxation time on the order of several picoseconds (See Fig.\ref{STFT2}($a$)), consistent with the relationship discussed. On the other hand, the other modes exhibit impulsive behavior (sine$-$like), and therefore the carrier relaxation time for the driving force coupled with these modes would be short lived, although more experimental and theoretical work is required to fully understand the different carrier dynamics in TMDC systems.


\begin{figure}
	\centering
	\includegraphics[width=8.8cm]{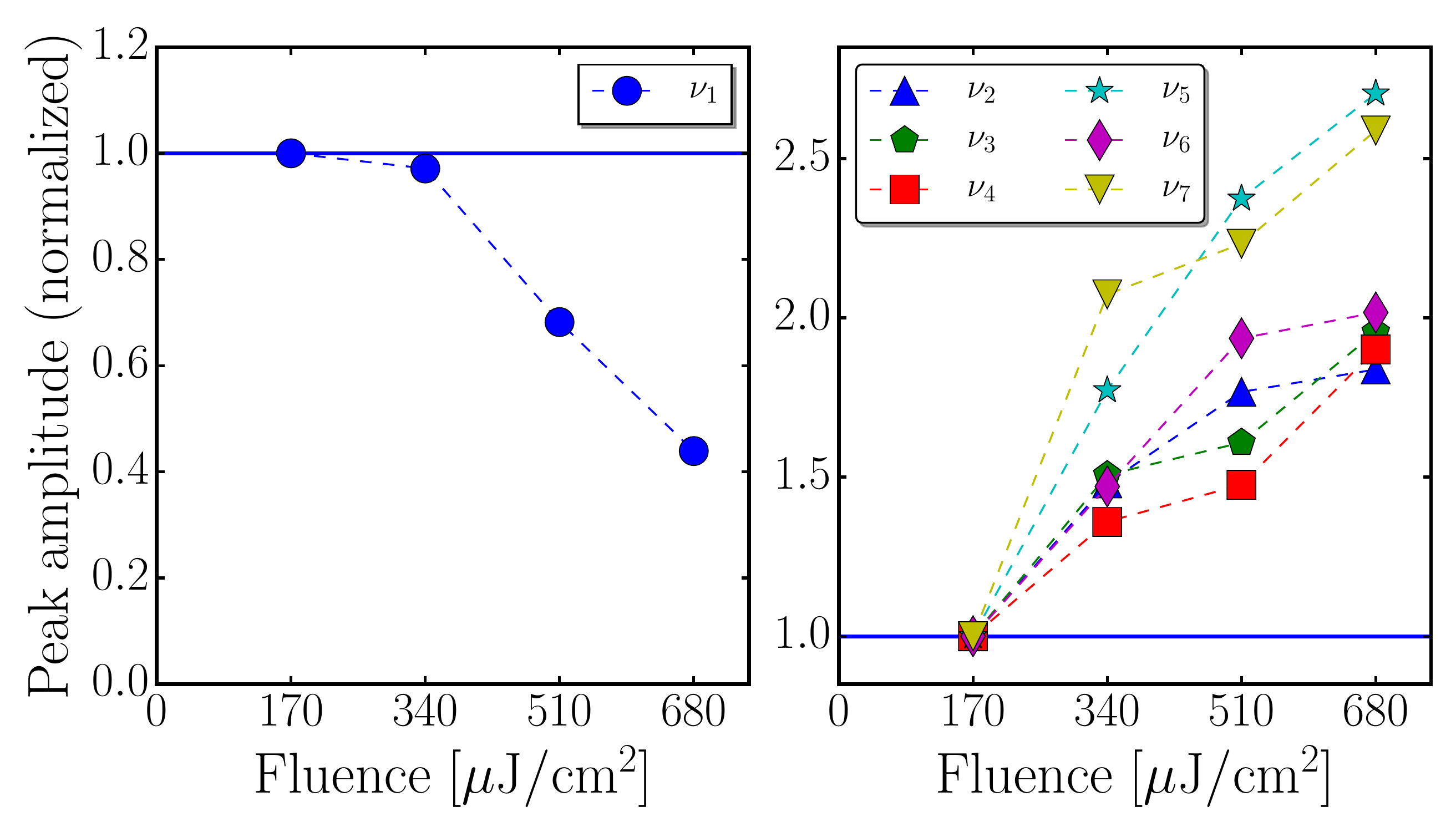}
	\caption{Fluence dependent peak amplitude shifts normalized by each of the values for $F=170$ $\mu \mathrm{J/cm^2}$ for $\nu_1$ ($a$) and $\nu_2 \sim \nu_7$ ($b$) obtained from the FT spectra displayed in  Fig.\ref{results}($d$). 
	}
	\label{peakinf}
\end{figure}

Fig.\ref{peakinf} shows the peak amplitudes versus the pump fluence obtained from Fig.\ref{results}($d$), being plotted for the low frequency $\nu_1$ mode in Fig.\ref{peakinf}($a$) and for the other optical modes in Fig.\ref{peakinf}($b$). First, for the $\nu_2 \sim \nu_7$ cases, Fig.\ref{peakinf}($b$) shows that the peak intensities (amplitudes) almost linearly increase with increasing pump fluence. This phenomenon can be understood as a characteristic property of CPs,
as observed for conventional semiconductors and metals \cite{hase2002dynamics,hase2015femtosecond,flock2014coherent}. In contrast, the peak intensity of the $\nu_1$ mode surprisingly decreased with increasing pump fluence, as seen in Fig.\ref{peakinf}($a$).
This behavior was opposite to that of the characteristic CP modes $\nu_2 \sim \nu_7$, suggesting that the $\nu_1$ mode is sensitive to the lattice temperature ($T_{l}$) \cite{chen2016activation,zhang2016raman}, as very recently reported by Zhang, \textit{et al}.\cite{zhang2019light}. 
Although the amplitude near time zero ($\xi_{1}$) increases with excitation fluence, the lattice temperature rise promotes faster dephasing of the coherent phonons, resulting in the decrease of FT amplitude as shown in Fig.\ref{peakinf}($a$) \cite{he2016coherent,hase1998PRB}.

Comparing the measurements taken at ultra$-$low temperature ($T = 4\ \mathrm{K}$) \cite{zhang2019light} and our present measurements at RT, the pump fluence dependence of the $T_{l}$ is expected to be significantly different. In fact, the maximum pump fluence used by Zhang \textit{et al.} was up to $\approx$ 5 $\mathrm{mJ/cm}^{2}$, while in the current experiments was only 680 $\mu\mathrm{J/cm}^{2}$, a fluence value one order less. Thus, the rise of $T_{l}$ in the current study is also expected to be one order of magnitude less than in the former study.
According to two$-$temperature model (TTM) calculations \cite{hohlfeld2000electron}, the maximum increase in the $T_{l}$ in the \tmote bulk crystal surface can be estimated to be from 305 K to 321 K, for pump fluences 170 to 680 $\mu\mathrm{J/cm}^{2}$, where the initial temperature was set as $T_{l}$ = 300\ $\mathrm{K}$. This TTM calculation result shows the pump pulse irradiation results in a rise of $\Delta T_{l}\approx 20\ \mathrm{K}$ at the sample surface. 
Although the phase transition between the 1T' and $\mathrm{T_d}$ phases has been believed to occur at 250 K, our data strongly suggest that the low frequency $\nu_1$ phonon in the $\mathrm{T_d}$ phase appears upon photoexcitation even at RT. A possible mechanism for the light$-$induced structural phase transition from 1T’ to $\mathrm{T_d}$ phases is the transient displacement of the nuclear equilibrium position of the $\nu_1$ mode, which is induced and stabilized by electron doping via photoexcitation, as has been recently proposed by Kim \textit{et al} \cite{kim2017PRB}. The pump$-$induced temperature rise ($\Delta T_{l}$) may partly contribute to the decrease in the peak amplitude [Fig.\ref{peakinf}($a$)] since higher temperature will destroy the $\mathrm{T_d}$ phase \cite{kim2017PRB}. 
The transient phase transition between the normal semimetal 1T' and the WSM T$_\mathrm{d}$ phases observed in the sub$-$picosecond time domain is expected to be a key phenomenon in controlling the process of the creation and annihilation of Weyl fermions and allows the development of ultrafast phase switching device applications at RT.

In conclusion, we observed a long lifetime low frequency shear phonon in the \tmote phase even at RT using pump$-$probe spectroscopy, which was thought to be present in only the $\mathrm{T_d}$ phase below 250 K.
The observed behavior implies the existence of a transient phase transition from the normal semimetal 1T' phase to the Type$-$II WSM T$_\mathrm{d}$ phase or a mixture between both phases in the ultrafast sub$-$picosecond time domain.
The SWFT analysis revealed that the amplitude of the shear phonon, which is considered to result in lattice symmetry breaking in the bulk crystal, is largely induced just after excitation, caused by a resonant ISRS or DECP mechanism.
High fluence laser irradiation leads the shear phonon mode to decrease in amplitude due to a lattice temperature rise of $\approx$ 20 K as calculated by a TTM analysis.
This study on ultrafast dynamics in the 1T' phase is in agreement with the existence of previously reported lattice symmetry switching between the normal semimetal to the WSM phase even at RT, and also suggests a new technique to unveil the sub$-$picosecond time domain creation and annihilation of Weyl fermions as well as their thermal dependence which is expected to lead to new device applications using topological phase switching.

This work was supported by JSPS KAKENHI (Grant Numbers. 17H02908 and 19H02619), and CREST, JST (Grant Number. JPMJCR1875), Japan.
We gratefully acknowledge R. Mondal for helping data analysis, and R. Ishikawa, T. Mori at Saitama University for measuring Raman spectrum data of our \tmote sample.

\bibliography{1TMoTe2_APL}

\end{document}